\documentclass[twocolumn,showpacs,preprintnumbers,amsmath,amssymb]{revtex4}

\usepackage{graphicx}
\usepackage{dcolumn}
\usepackage{bm}

\usepackage[english]{babel}
\usepackage{amsmath}

\begin{document}

\preprint{APS/123-QED}

\title{ Flexible Logic from Neuronal Dynamics  }

\author{Abraham Miliotis }
\email{aris.miliotis@gmail.com}
\author{ Sachin Talathi }
\email{sachin.talathi@gmail.com}
\author{William Ditto    }
\email{william.ditto@bme.ufl.edu}
\affiliation{%
J Crayton Pruitt Department of Biomedical Engineering, University of Florida, Gainesville, FL  32611\\
}%

\date{\today}

\begin{abstract}
We present two novel methods for performing logic operations.\ Our methods are based on using the time dimension for programming and data representation. The first method is based on varying the sampling moment in time of a neuronal action potential, and the second method is based on a neural delay system, where the generation of the action potential is delayed by specific time lengths, to be sampled at a fixed moment in time. Both methods are supported by explicit examples.\end{abstract}

\pacs{Valid PACS appear here}
\maketitle

The computational capabilities of chaotic and non-linear systems have been widely reported \cite{sinha:2156,sinha:363,sinha:036216,sinha:036214,murali:025201,murali:2005,murali:2669,murali:016205,munakata:1629}. Most of these computational methods involve the threshold control of a chaotic system \cite{murali:016210,ding:644} to perform computations, whether its simple arithmetic calculations \cite{sinha:363,sinha:2156}, emulation of logic gates \cite{sinha:036216} or solving more complex computational problems like the Deutsch-Jozsa problem \cite{sinha:036214}. These nonlinear systems, with variable thresholding schemes, provide  an unique approach to emulate all logic gates and have the flexibility of switching between different operational roles, thus allowing for the design of a dynamic computer architecture. More recently a different method for logic gate emulation  based on the synchronization of a driver and response nonlinear systems has been reported \cite{murali:025201}. In this method though, the programming instruction and input data are separated in different parts of the system. In this paper we present two novel  methods to perform computation using nonlinear systems which utilize time as computational commands to represent both the programming instruction and the input data stream into the logic gate, i.e. computation is performed by varying a single parameter. The first method is based on variation of the observation time (sampling instance) of a non-linear signal to obtain logic gate emulation, while the second method for logic gate emulation is based on time delays in the generation of a non-linear signal.

We will demonstrate these two general methods for logic gate emulation by using the nonlinear properties of action potentials generated by neurons. We model neuronal dynamics in the framework of conductance based Hodgkin-Huxley (HH) neurons \cite{Talathi,Talathi2}. The first method for logic emulation is based on the idea of sampling the membrane voltage signal of a single neuron at different moments in time.  The second method utilizes two bi-directionally coupled HH neurons, such that the two neuron system creates a time delay circuit. This two neuron time delay circuit operates by generating an output spike at time $t_{0}+\tau(R)$ in response to an input spike arriving into the circuit at time $t_{0}$. The logic gates are emulated by varying the synaptic strength $R$ that determines the time delay $\tau(R)$ and then observing the output at a predetermined fixed time instant.

A system (a flexible logic gate) to be able to switch between the five fundamental logic gates (AND, NAND, OR, NOR, XOR), needs to be able to reproduce the truth table of each and every one of these gates \cite{murali:025201,sinha:036216}. We can combine the truth tables into a single non-linear function of the form: $F(u,t) > 0, \text{for } t-\Delta t >t>t+\Delta\ t,\text{ else } F(u,t) =0$; which provides the required behavior for a system to be utilized as a flexible logic gate.
This function is very similar to an action potential generated by a neuron; resting at a low voltage ($F(u,t)=0$), and for a brief length of time rising very rapidly to a higher voltage ($F(u,t)>0$), and then dropping very rapidly back to its original level. It is known that neurons have a method for performing computational operations \cite{Zhang}; we do not propose that the methods we present here are the method neurons use, but it may well be a way by which neuronal networks in the brain communicate information.

\begin{figure}
\includegraphics[scale=0.75]{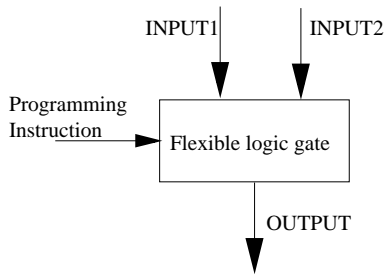}
\caption{\label{fig:flexgate} Required inputs and outputs of a flexible logic gate.}
\end{figure}

For any computational system to be able to perform flexible logic there are three parameters that need to be given to the system. The first parameter that needs to be introduced to the system is the programming instruction. This programming instruction is the parameter of the system that defines which of the five logic gates will be performed on a given set of inputs. The other two parameters that need to be given to the system are the two logical inputs, INPUT1 and INPUT2, see Figure \ref{fig:flexgate}.

\begin{figure}
\includegraphics[scale=0.2]{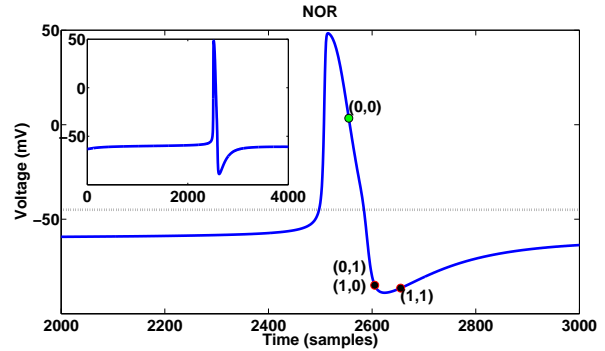}
\caption{\label{fig:stNOR} Demonstration of operating a NOR gate using different sampling times. Green dot represents a logical 1 at the OUTPUT, blue dot represents logical 0 at the OUTPUT. The three dots indicate the distinct moments in time that the signal would be sampled to generate the truth table of a NOR gate. Inset: A typical action potential.}
\end{figure}

The first method of flexible logic implementation we will introduce is based on utilizing fixed time intervals for representing the programming instruction and the two inputs to the gate, as described above. These time intervals are combined (as explained below) to determine the time at which the signal is to be sampled to obtain an output from the gate. Consider a periodic nonlinear signal of period T, which has a form close to that of a neuronal action potential, see inset of Figure~\ref{fig:stNOR}, i.e. for most of the time the signal is ``low'' and for a brief length of time the signal is ``high''. A programming instruction is a time length, $t_{prog}<T$, at which the signal is sampled; this time length is measured from a predetermined reference time point. Next, the input data are also represented by a predetermined fixed time length. Specifically  a finite nonzero time length represents a logical 1, $t_{input,i}$, where $i=\{1,2\}$, while zero time length representing a logical 0. So we have:
\begin{displaymath}
t_{input,i} = \{\begin{array}{cc}
t_{input,i}>0, & INPUTi=1 \\ \\
t_{input,i}=0, & INPUTi=0  \\ 
\end{array}
\end{displaymath}
These three time intervals define the observation time at which the signal is sampled for the output of the logic gate. This observation time is given by $t_{prog}+t_{input,1}+t_{input,2}$.

 It is important to note that the input data time length needs to be a constant length irrespective of which of the two inputs, INPUT1 or INPUT2, is at logical 1 and of double time length for the input (INPUT1, INPUT2) =(1,1). This is a necessary condition so that the two input streams can be considered degenerate, i.e. the cases (0,1) and (1,0) are represented by an equal time length, while the case (1,1) corresponds to double this time length, which we call as the input ``unit'' time. Careful choice of both the programming instruction time and the input unit time can produce responses (threshold crossing) at the sampled instance that are identical to the fundamental logic gates. Note that since the time length representing a logical 0 is of zero length, the programming instruction time is analogous to performing the logical operation between two inputs of logical 0, i.e. (0,0).
 

As an example consider the action potential generated by a single neuron, modeled as a type I HH neuron \cite{Talathi}, see inset of Figure~\ref{fig:stNOR}. For most of its period the signal rests at -60mV, the resting membrane potential of the neuron, but for a brief period of time the membrane voltage rises above the resting potential when the neuron generates an action potential. If we interpret a membrane voltage over -45mV as a logical 1 and below as logical 0, then it is only a matter of $when$ we observe the signal to obtain a logical 0 or a logical 1 at the output, which represents a flexible logic opearation. Take for example the case of the logical gate NOR, see Table \ref{tab:NORgate}, the truth table of the NOR gate is: OUTPUT=1 for INPUTS=(0,0), OUTPUT=0 for INPUTS=(0,1) / (1,0) and OUTPUT=0 for INPUTS=(1,1). So we need 3 distinct times at which to observe(sample) our signal that correspond to the OUTPUT values given by the truth table. At the same time, the time difference between the 3 observation times should be constant for the logical input of 1 to have a constant representation, irrespective if it is INPUT1 or INPUT2. See Table \ref{tab:NORgate}.
\begin{table}
\caption{\label{tab:NORgate} NOR gate truth table, sampling time representation of programming and inputs, necessary conditions for representation of the NOR gate.}
\begin{ruledtabular}
\begin{tabular}{c|ccccc}
  INPUTS        & NOR & Sampling time                   & Condition     \\
 (0,0)          & 1   & $t_{prog}+0+0$                    &$V_{sampled}>-45mV$               \\
 (0,1)/(1,0)    & 0   & $t_{prog}+t_{input,1/2}+0$         &$V_{sampled}<-45mV$               \\
 (1,1)          & 0   & $t_{prog}+t_{input,1}+t_{input,2}$  &$V_{sampled}<-45mV$               \\
\end{tabular}
\end{ruledtabular}
\end{table}

In Figure~\ref{fig:stNOR} we show the results of sampling our signal at the time instances of 2555 for inputs (0,0), 2605 for inputs (0,1)/(1,0) and at 2655 for inputs (1,1). Therefore in terms of programming and input time lengths, for the NOR gate, $t_{prog} = 2555$ time steps and $t_{input,i}=50, i=\{1,2\}$ time steps. As is clear from the figure the appropriate OUTPUT values are obtained, i.e. the voltage is over -45mV for the (0,0) case and below for the other two cases. This observation can be interpreted as follows: to perform the logic gate NOR we require to wait 2555 time steps as a programming time length and then another 50 time steps for each occurrence of a single logical input of 1. Using the same system, that is the HH neuron described above with the action potential generated at the same rate, time instances for accomplishing the all fundamental gates are given in Table~\ref{tab:stGates}. So in essence we have two time delays on the observation instance, one for programming instruction, i.e. which gate will be performed, and it is analogous to performing an operation on inputs (0,0), and second time delay representing an input of logical 1, in this case it is equal to 50 time steps.

\begin{table}
\caption{\label{tab:stGates} Appropriate time sample instances to perform each fundamental logic gate, for each case of different inputs. In brackets the output of each gate is given.}
\begin{ruledtabular}
\begin{tabular}{c|ccccc}
                & NOR     & NAND    & AND     & OR      &  XOR  \\
 (0,0)          & 2555 (1) & 2505 (1) & 2405 (0) & 2455 (0) & 2485 (0)\\
 (0,1)/(1,0)    & 2605 (0) & 2555 (1) & 2455 (0) & 2505 (1) & 2535 (1)\\
 (1,1)          & 2655 (0) & 2605 (0) & 2505 (1) & 2555 (1) & 2585 (0)\\
\end{tabular}
\end{ruledtabular}
\end{table}

The second method for flexible logic implementation  is based on the idea that one fixes the time instance of observation and varies the time of generation of the action potential to perform the  logic operations. This form of variable delays can be implemented, in a neuronal system, by using a simple network of two mutually coupled neurons as explained below.

\begin{figure}
\includegraphics[scale=0.4]{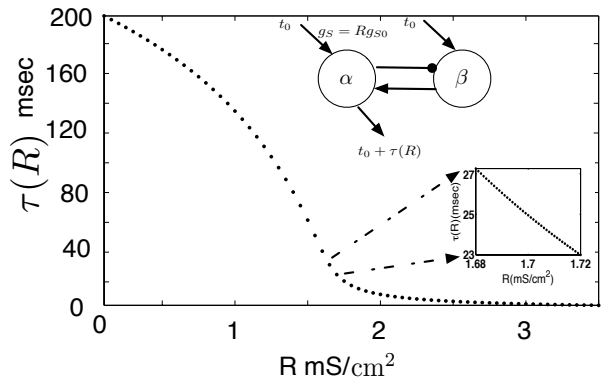}
\caption{\label{fig:DelayModel} Two neuron time delay circuitry. (Adapted from \cite{Talathi}.) }
\end{figure}

In \cite{Talathi} the authors present a neuronal circuit that has the ability to generate an action potential at a delayed time interval controlled by synaptic coupling strength. In brief, the circuit is composed of two HH neurons arranged as shown in Figure \ref{fig:DelayModel} inset; with neuron ($\alpha$), set at resting state and the bistable neuron $\beta$ also set at resting fixed point state. When an action potential arrives at time $t_{0}$, the neuron $\beta$ is pushed into its bistable oscillating state. This neuron then sends an excitatory input drive to neuron $\alpha$, which eventually triggers an action potential, at a delayed time interval $t_{0}+\tau(R)$, that depends on the strength of excitatory synaptic input the neuron receives at time $t_{0}$ through the synapse $g_{S}=Rg_{S0}$, with $g_{S0}=1$. At that stage the action potential generated by the neuron $\alpha$ will inhibit the bistable neuron sending it back to its resting state. Thus both neurons will return to their resting states, making the system receptive to a new operation. The time delay for the generation of the action potential is governed by the parameter $R$, as can be seen from Figure \ref{fig:DelayModel}. By varying $R$ we can have the generated action potential be produced at different times with respect to an initiating spike. 

Now in order to implement flexible logic gates with this two neuron time delay circuitry, we set a specific observation time to observe the output of this circuitry in response to an initiating spike; i.e. in the example below this time is at 425msec after the initiating spike. Next we set a confirmation voltage at -45mV; which we interpret as: logical 1 output if voltage at observation time exceeds -45mV, else logical output 0. We also use the range of $R$, $1.68 < R < 1.72$ where the curve of Figure \ref{fig:DelayModel} is linear, so that the changes in $R$ are linearly proportional to the changes in $\tau(R)$ and so changes for a logical 1 at the input are independent of whether it is at INPUT1 or INPUT2.

Using this method of time delays, flexible logic can be implemented as follows: the time delay circuitry is setup with a specific $R_{prog}$ representing which logical operation will be performed. Further, we shift the total $R$ value of the system an extra amount depending on the inputs to the system. In analogy to the previous method we have the total value of $R$ defined as: $R = R_{prog}+R_{input,1}+R_{input,2}$; representing the combination of programming instruction, which decides the logic operation to be performed, and the two input data streams, each represented by a shift in total $R$. In our specific example a shift of $R_{input,i}=0.005,i=\{1,2\}$ represents a single logical 1 at the inputs, for cases (0,1) and (1,0); and naturally for inputs (1,1) $R_{input,1}+R_{input,2}=0.01$ and for (0,0) $R_{input,i}=0,i=\{1,2\}$. 

\begin{figure}
\includegraphics[scale=0.2]{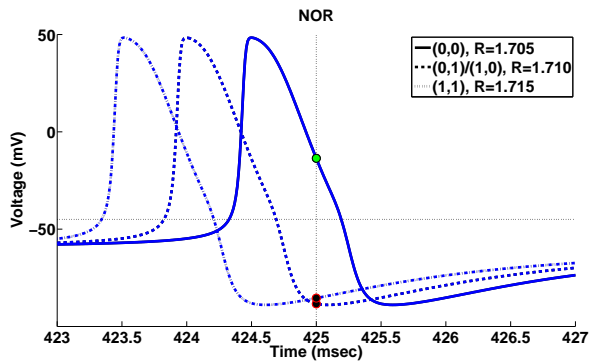}
\caption{\label{fig:dtNOR} Demonstration of operating a NOR gate using different delay times. Green dot represents a logical 1 at the OUTPUT, blue dot represents logical 0 at the OUTPUT. The three action potentials each represents a distinct case of the truth table of NOR gate, in reality only one action potential will be generated for the specific case of inputs.}
\end{figure}

An initiating spike is given to the system every second, like a universal clock. At 425msec after the initiating spike we observe the system, if there is an action potential and the voltage is higher than -45mV we interpret a logical 1 at the output otherwise a logical 0, see Figure~\ref{fig:dtNOR} for an illustrative example of a NOR gate implementation. In Figure \ref{fig:dtNOR} we see the three distinct cases of the truth table of a NOR gate superimposed. Each case is generated with a different $R$ value, representing the programming and the inputs to the gate. As is expected for the NOR gate, only in the case of INPUTS=(0,0) we have an action potential, at the observation time, over -45mV, signified by the green dot. In the other two cases of the truth table the action potentials generated at those $R$ values are lower than -45mV at the observation time, signified by blue dots. Just like in the previous method, of varying sampling time, we can with this method reproduce the five fundamental logic gates with different delay parameter values, $R$, see Table \ref{tab:dtGates}.

\begin{table}
\caption{\label{tab:dtGates} R values for all gates.}
\begin{ruledtabular}
\begin{tabular}{c|ccccc}
      INPUTS    & NOR    & NAND   & AND    & OR     &  XOR  \\
 (0,0)          & 1.705  & 1.700  & 1.690  & 1.695  & 1.697 \\  
 (0,1)/(1,0)    & 1.710  & 1.705  & 1.695  & 1.700  & 1.702 \\
 (1,1)          & 1.715  & 1.710  & 1.700  & 1.705  & 1.707 \\
\end{tabular}
\end{ruledtabular}
\end{table}

We are using neuronal systems for our demonstrations as they are one of the most natural generators for the function that covers all fundamental logic gates. The neuronal action potential structure is exactly what is needed for our methods to work. The novelty of the methods introduced is that we use time for both the computational programming and the data representation. Based on these methods our computational efficiency and capabilities are limited by how finely time can be sliced, the sampling rate. The finer definition we have on the slicing of time, higher sampling rate, the more distinct the different cases can be and more robust to noise. In the implementation of our ideas using neuronal models, the key parameter that defines the precision of each operation is the width of the action potential in relation to the period of the signal, the wider the action potential is the further apart in time each case will be providing more resolution between the different logic cases and more robustness to noise.

\begin{figure*}[h]
\includegraphics[scale=0.5]{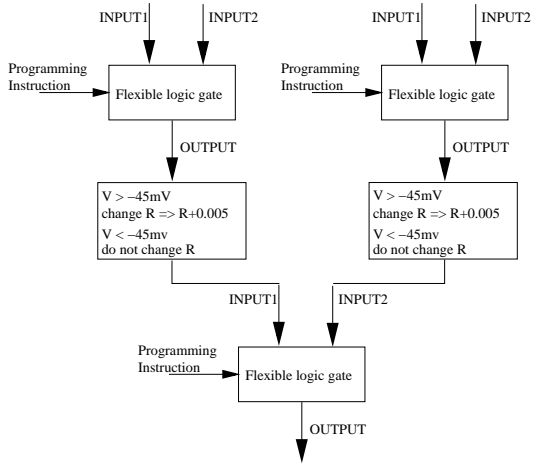}
\caption{\label{fig:concat} Each flexible logic gate passes its output to a look up table that translates the event of crossing the -45mV threshold to a change in the R parameter of the next flexible logic gate.}
\end{figure*}

To concatenate such systems into more complex logic circuits, of two logic gates and more, the output from one such system needs to be given as an input to the next gate (system). This can be accomplished with the use of a lookup table that relates the event of crossing the threshold, or not, with a time length (for the first method) or with a shift in $R$ (for the second method), see Figure \ref{fig:concat} for a demonstration. A look up table is used because the nature of the inputs to the system is different to that of the output, inputs are time lengths (or changes in synaptic strength, $R$) whereas outputs are events. 

Further research will focus to bring the inputs and outputs to the same units so that concatenation can be performed without the use of a lookup table, which adds computational overhead. In addition a method using time as computational commands, to store and process (specifically: searching) information will be reported in a future paper.


\bibliographystyle{plain}
\bibliography{CompWithTime}

\end{document}